\documentclass{article}
\usepackage{graphicx}

\begin{document}

\hspace{-0.6cm}
{\Large \bf Lattice QCD Study for Confinement and Hadrons}

\vspace{0.5cm}

\begin{center}
Hideo Suganuma\footnote{Invited Talk presented by Hideo Suganuma at 
the International Conference on 
``High Energy and Mathematical Physics" (ICHEMP05),  
4-7 April 2005, Cadi Ayyad University, Marrakech, Morocco.}, 
Hiroko Ichie 

\vspace{0.3cm}
{\it Faculty of Science, Tokyo Institute of Technology, \\
2-12-1 Ohokayama, Meguro, Tokyo 152-8551, Japan}\\
\vspace{0.5cm}

Fumiko Okiharu 

\vspace{0.3cm}
{\it Department of Physics, Nihon University, \\
1-8-14 Kanda Surugadai, Chiyoda, Tokyo 101-8308, Japan}\\
\vspace{0.5cm}

Toru T. Takahashi

\vspace{0.3cm}
{\it Yukawa Institute for Theoretical Physics, Kyoto University, \\
Kitashirakawa, Sakyo, Kyoto 606-8502, Japan}
\end{center}

\vspace{0.5cm}

\noindent
Using SU(3) lattice QCD, we perform the detailed studies of the three-quark and the multi-quark potentials. 
From the accurate calculation for 
more than 300 different patterns of 3Q systems, the static ground-state 3Q potential 
$V_{\rm 3Q}^{\rm g.s.}$ is found to be well described 
by the Coulomb plus Y-type linear potential (Y-Ansatz) within 1\%-level deviation.
As a clear evidence for Y-Ansatz, 
Y-type flux-tube formation is actually observed on the lattice in maximally-Abelian projected QCD.
For about 100 patterns of 3Q systems, 
we perform the accurate calculation for the 1st excited-state 3Q potential $V_{\rm 3Q}^{\rm e.s.}$
by diagonalizing the QCD Hamiltonian in the presence of three quarks, and find 
a large gluonic-excitation energy $\Delta E_{\rm 3Q} \equiv V_{\rm 3Q}^{\rm e.s.}-V_{\rm 3Q}^{\rm g.s.}$ 
of about 1 GeV, which gives a physical reason of  
the success of the quark model.
$\Delta E_{\rm 3Q}$ is found to be reproduced by the ``inverse Mercedes Ansatz'', 
which indicates a complicated bulk excitation for the gluonic-excitation mode. 
We study also the tetra-quark and the penta-quark potentials in lattice QCD, 
and find that they are well described by 
the OGE Coulomb plus multi-Y type linear potential, which supports the flux-tube picture even for the multi-quarks.

\section{Introduction}

In 1966, Yoichiro Nambu \cite{N66} first proposed the SU(3) gauge theory, i.e., quantum chromodynamics (QCD),  
as a candidate for the fundamental theory of the strong interaction, 
just after the introduction of the ``new" quantum number, ``color" \cite{HN65}.
In 1973, the asymptotic freedom of QCD was theoretically proven \cite{GWP73}, 
and, through the applicability check of perturbative QCD to high-energy hadron reactions, 
QCD has been established as the fundamental theory of the strong interaction.
Even at present, however, it is very difficult to deal with QCD analytically due to its strong-coupling nature in the infrared region.
Indeed, in spite of its simple form, QCD creates thousands of hadrons and leads to various interesting nonperturbative phenomena 
such as color confinement \cite{N6970,N74,conf2003} and dynamical chiral-symmetry breaking \cite{NJL61}.
Instead, lattice QCD has been applied as the direct numerical analysis for nonperturbative QCD.

In 1979, the first application \cite{C7980} of lattice QCD Monte Carlo simulations was done 
for the inter-quark potential between a quark and an antiquark using the Wilson loop.
Since then, the study of the inter-quark force has been one of the important issues in lattice QCD \cite{R97}.
Actually, in hadron physics, the inter-quark force can be regarded as an elementary quantity 
to connect the ``quark world" to the ``hadron world", and plays an important role to hadron properties. 

In this paper, we perform the detailed and high-precision analyses 
for the inter-quark forces in the three-quark and the multi-quark systems with SU(3) 
lattice QCD \cite{TMNS99,TMNS01,TSNM02,TS03,TS04,STI04,STOI04,OST04,OST04p,SOTI04,OST05}, 
and try to extract the proper picture of hadrons.

\section{The Three-Quark Potential in Lattice QCD}

In general, the three-body force is regarded as a residual interaction in most fields in physics.
In QCD, however, the three-body force among three quarks is 
a ``primary" force reflecting the SU(3) gauge symmetry.
In fact, the three-quark (3Q) potential is directly responsible 
for the structure and properties of baryons, 
similar to the relevant role of the Q$\bar{\rm Q}$ potential for meson properties. 
Furthermore, the 3Q potential is the key quantity to clarify the quark confinement in baryons.
However, in contrast to the Q$\bar{\rm Q}$ potential \cite{R97},
there were almost no lattice QCD studies  
for the 3Q potential before our study in 1999 \cite{TMNS99},
in spite of its importance in hadron physics. 

As for the functional form of the inter-quark potential, we note two theoretical arguments 
at short and long distance limits.
\begin{itemize}
\item[1.]
At the short distance, perturbative QCD is applicable,
and therefore inter-quark potential is expressed as the sum of the two-body OGE Coulomb potential. 
\item[2.]
At the long distance, the strong-coupling expansion of QCD is plausible, and it 
leads to the flux-tube picture \cite{KS75CKP83}.
\end{itemize}
Then, we theoretically conjecture the functional form of the inter-quark potential 
as the sum of OGE Coulomb potentials and the linear potential based on the flux-tube picture.
Of course, it is highly nontrivial that these simple arguments on UV and IR limits of QCD hold for the intermediate region. 
Nevertheless, the Q$\bar {\rm Q}$ potential $V_{\rm Q\bar Q}(r)$ is well described with this form 
as \cite{R97,TMNS01,TSNM02} 
\begin{eqnarray}
V_{\rm Q \bar Q}(r)=-\frac{A_{\rm Q\bar Q}}{r}+\sigma_{\rm Q \bar Q}r+C_{\rm Q\bar Q}.
\label{VQQ}
\end{eqnarray}
For the 3Q system, there appears a junction which connects the three flux-tubes from the three quarks, 
and Y-type flux-tube formation is expected \cite{TMNS01,TSNM02,KS75CKP83,FRS91,BPV95}.
Therefore, the (ground-state) 3Q potential is expected to be 
the Coulomb plus Y-type linear potential, i.e., Y-Ansatz,
\begin{eqnarray}
V_{\rm 3Q}^{\rm g.s.}=-A_{\rm 3Q}\sum_{i<j}\frac1{|{\bf r}_i-{\bf r}_j|}+
\sigma_{\rm 3Q}L_{\rm min}+C_{\rm 3Q},
\label{V3Q}
\end{eqnarray}
where $L_{\rm min}$ is the length of the Y-shaped flux-tube.

For more than 300 different patterns of spatially-fixed 3Q systems, 
we calculate the ground-state 3Q potential $V_{\rm 3Q}^{\rm g.s.}$ 
from the 3Q Wilson loop $W_{\rm 3Q}$ 
using SU(3) lattice QCD \cite{TMNS01,TSNM02,TS03,TS04} 
with the standard plaquette action at the quenched level 
on various lattices, i.e.,  
($\beta$=5.7, $12^3\times 24$),
($\beta$=5.8, $16^3\times 32$), 
($\beta$=6.0, $16^3\times 32$) and 
($\beta=6.2$, $24^4$).
For the accurate measurement, we construct the ground-state-dominant  
3Q operator using the smearing method.
Note that the lattice QCD calculation is completely independent of any Ansatz for the potential form.

To conclude, we find that the static ground-state 3Q potential $V_{\rm 3Q}^{\rm g.s.}$
is well described by the Coulomb plus Y-type linear potential (Y-Ansatz)  
within 1\%-level deviation \cite{TMNS01,TSNM02}.
To demonstrate this, we show in Fig.1 the 3Q confinement potential $V_{\rm 3Q}^{\rm conf}$, 
i.e., the 3Q potential subtracted by the Coulomb part, 
plotted against the Y-shaped flux-tube length $L_{\rm min}$.
For each $\beta$, clear linear correspondence is found between the 3Q confinement potential 
$V_{\rm 3Q}^{\rm conf}$ and $L_{\rm min}$, 
which indicates Y-Ansatz for the 3Q potential. 

\begin{figure}[h]
\begin{center}
\includegraphics[height=6.5cm]{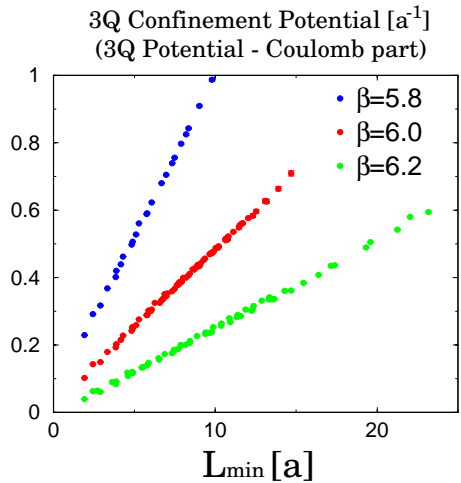}
\caption{
The 3Q confinement potential $V_{\rm 3Q}^{\rm conf}$, 
i.e., the 3Q potential subtracted by the Coulomb part, 
plotted against 
the total flux-tube length $L_{\rm min}$ of Y-Ansatz
in the lattice unit.
}
\end{center}
\end{figure}

\begin{figure}[h]
\begin{center}
\includegraphics[height=4.8cm]{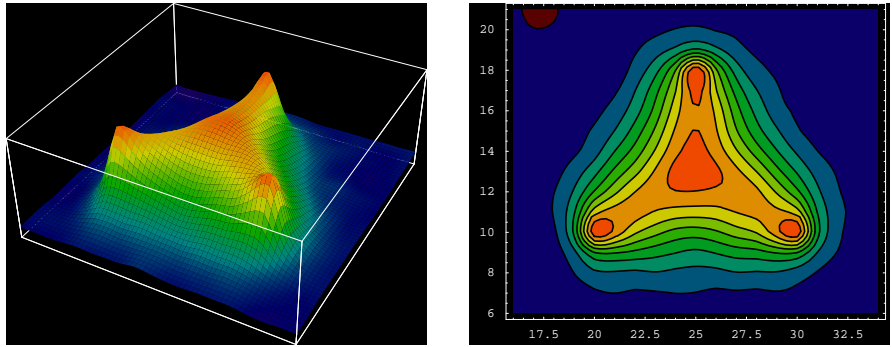}
\caption{
The lattice QCD result for Y-type flux-tube formation 
in the spatially-fixed 3Q system 
in MA projected QCD.
The distance between the junction and each quark is about 0.5 fm.
}
\end{center}
\end{figure}

Recently, as a clear evidence for Y-Ansatz, 
Y-type flux-tube formation is actually observed 
in maximally-Abelian (MA) projected lattice QCD 
from the measurement of the action density  
in the spatially-fixed 3Q system \cite{STI04,STOI04,SOTI04,IBSS03}. (See Fig.2.) 
In this way, together with recent several analytical studies \cite{KS03,C0405} 
and other lattice-QCD work \cite{BS04}, Y-Ansatz for the static 3Q potential seems to be almost settled. 

\section{The Gluonic Excitation in the 3Q System}

In this section, 
we study the excited-state 3Q potential and the gluonic excitation in the 3Q system 
using lattice QCD \cite{TS03,TS04}.
The excited-state 3Q potential $V_{\rm 3Q}^{\rm e.s.}$ is 
the energy of the excited state in the static 3Q system.
The energy difference $\Delta E_{\rm 3Q} \equiv V_{\rm 3Q}^{\rm e.s.}-V_{\rm 3Q}^{\rm g.s.}$ 
between $V_{\rm 3Q}^{\rm g.s.}$ and $V_{\rm 3Q}^{\rm e.s.}$ 
is called as the gluonic-excitation energy, and 
physically means the excitation energy of the gluon-field configuration 
in the static 3Q system.
In hadron physics, the gluonic excitation \cite{TS03,TS04,JKM03} 
is one of the interesting phenomena 
beyond the quark model, and relates to the hybrid hadrons such as $q\bar qG$ 
and $qqqG$ in the valence picture. 

\begin{figure}[h]
\begin{center}
\includegraphics[height=4.3cm]{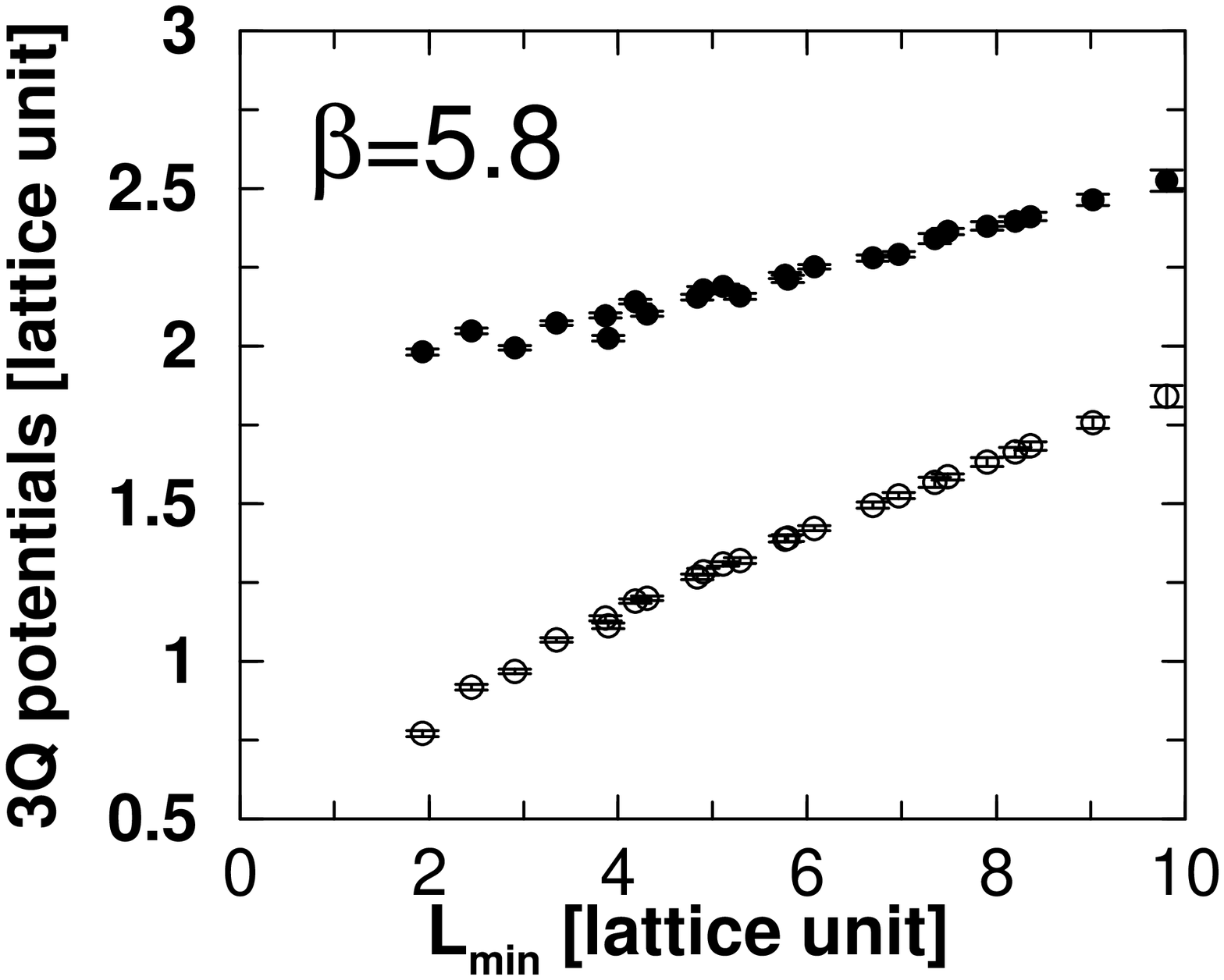}
\includegraphics[height=4.3cm]{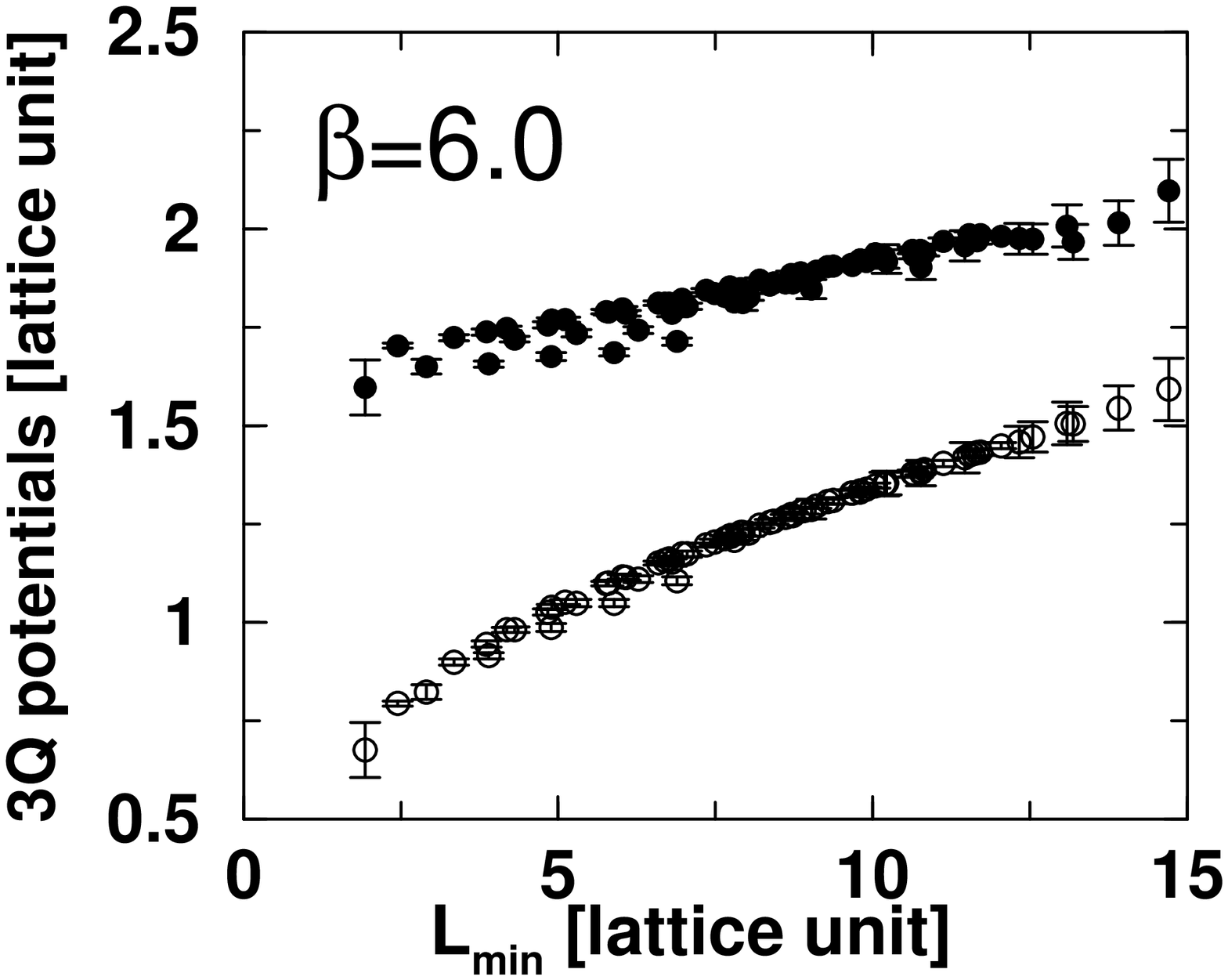}
\caption{
The 1st excited-state 3Q potential 
$V_{\rm 3Q}^{\rm e.s.}$ and 
the ground-state 3Q potential $V_{\rm 3Q}^{\rm g.s.}$.
The lattice results at $\beta=5.8$ and $\beta=6.0$ well coincide 
except for an irrelevant overall constant. 
The gluonic excitation energy 
$\Delta E_{\rm 3Q} \equiv V_{\rm 3Q}^{\rm e.s.}-V_{\rm 3Q}^{\rm g.s.}$ 
is about 1GeV in the hadronic scale 
as $0.5{\rm fm} \le L_{\rm min} \le 1.5{\rm fm}$.
}
\end{center}
\vspace{-0.5cm}
\end{figure}

For about 100 different patterns of 3Q systems, 
we calculate the excited-state potential 
in SU(3) lattice QCD with $16^3\times 32$ at $\beta$=5.8 and 6.0 at the quenched level 
by diagonalizing the QCD Hamiltonian in the presence of three quarks. 
In Fig.3, we show the 1st excited-state 3Q potential $V_{\rm 3Q}^{\rm e.s.}$ and 
the ground-state potential $V_{\rm 3Q}^{\rm g.s.}$.
The gluonic excitation energy $\Delta E_{\rm 3Q} \equiv V_{\rm 3Q}^{\rm e.s.}-V_{\rm 3Q}^{\rm g.s.}$ 
in the 3Q system is found to be about 1GeV 
in the hadronic scale as $0.5{\rm fm} \le L_{\rm min} \le1.5{\rm fm}$.
Note that the gluonic excitation energy of about 1GeV is rather large compared with 
the excitation energies of the quark origin. 
This result predicts that the lowest hybrid baryon $qqqG$ has a large mass of about 2 GeV.

\subsection{Inverse Mercedes Ansatz for the Gluonic Excitation}

Next, we investigate the functional form of the gluonic excitation energy in 3Q systems \cite{TS04,C0405},
$\Delta E_{\rm 3Q} \equiv V_{\rm 3Q}^{\rm e.s.}-V_{\rm 3Q}^{\rm g.s.}$, 
where the Coulomb part is expected to be canceled between $V_{\rm 3Q}^{\rm g.s.}$ and $V_{\rm 3Q}^{\rm e.s.}$. 
After some trials, as shown in Fig.4, we find that the lattice data of the gluonic excitation energy 
$\Delta E_{\rm 3Q} \equiv V_{\rm 3Q}^{\rm e.s.}-V_{\rm 3Q}^{\rm g.s.}$ 
are relatively well reproduced by the ``inverse Mercedes Ansatz'' \cite{TS04}, 
$\Delta E_{\rm 3Q} =\frac{K}{L_{\rm\bar Y}}+G, \quad
L_{\rm\bar Y} \equiv {\sum_{i=1}^{3}\sqrt{x_i^2-\xi x_i+\xi^2}}
=\frac12 \sum_{i\ne j}\overline{{\rm P}_i{\rm Q}_j}
\quad
(x_i \equiv \overline{\rm PQ}_i, \ \xi \equiv \overline{\rm PP}_i)$, 
where $L_{\rm \bar Y}$ denotes the ``modified Y-length" defined by the half perimeter of the ``Mercedes form"  
as shown in Fig.4(a).
As for ($K$, $G$, $\xi$), we find 
($K \simeq 1.43$, $G\simeq$ 0.77 GeV, $\xi \simeq$ 0.116 fm) at $\beta=5.8$, and 
($K \simeq 1.35$, $G \simeq$ 0.85 GeV, $\xi \simeq$ 0.103 fm) at $\beta=6.0$.

\begin{figure}[h]
\begin{center}
\includegraphics[height=3.3cm]{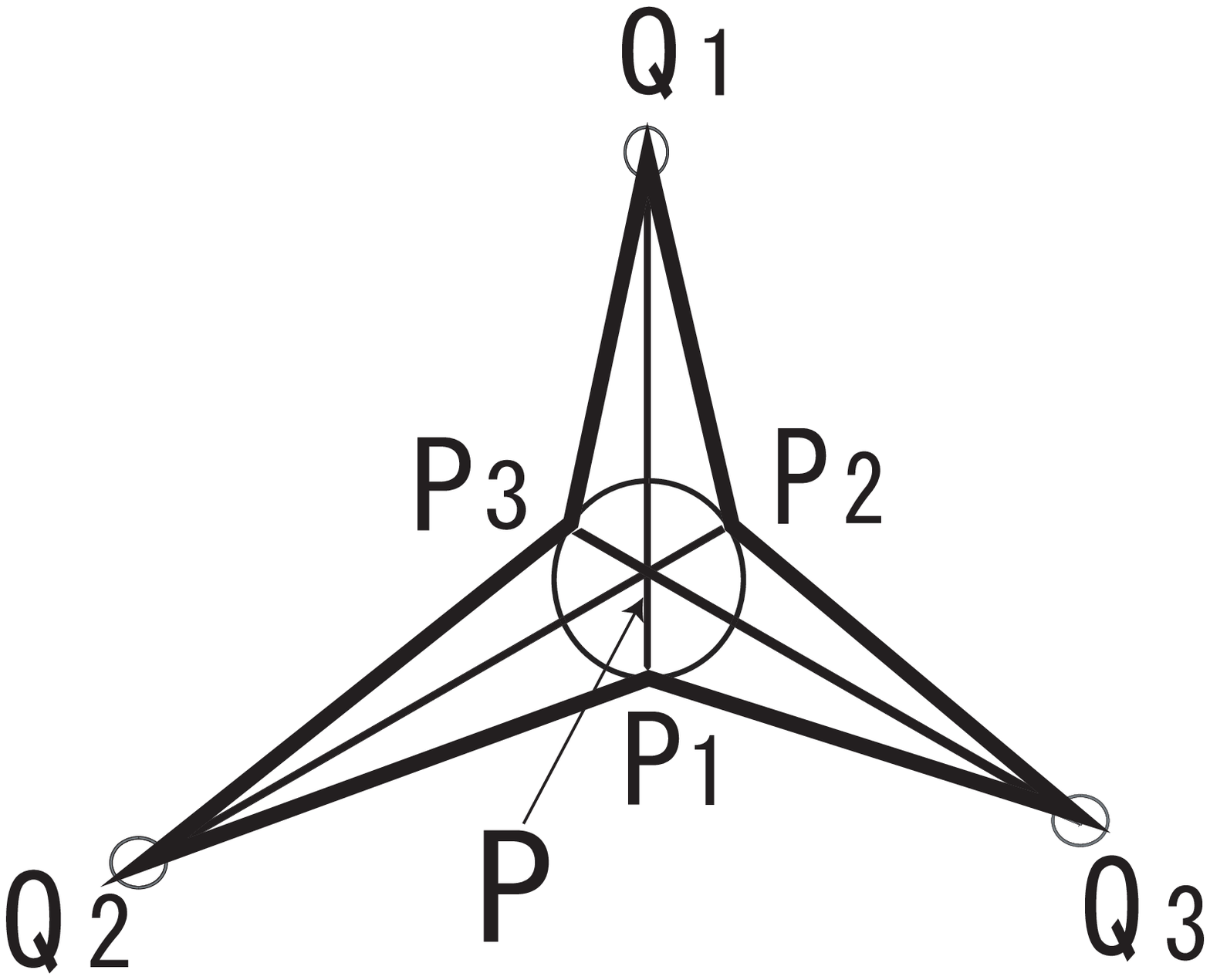}
\includegraphics[height=3.2cm]{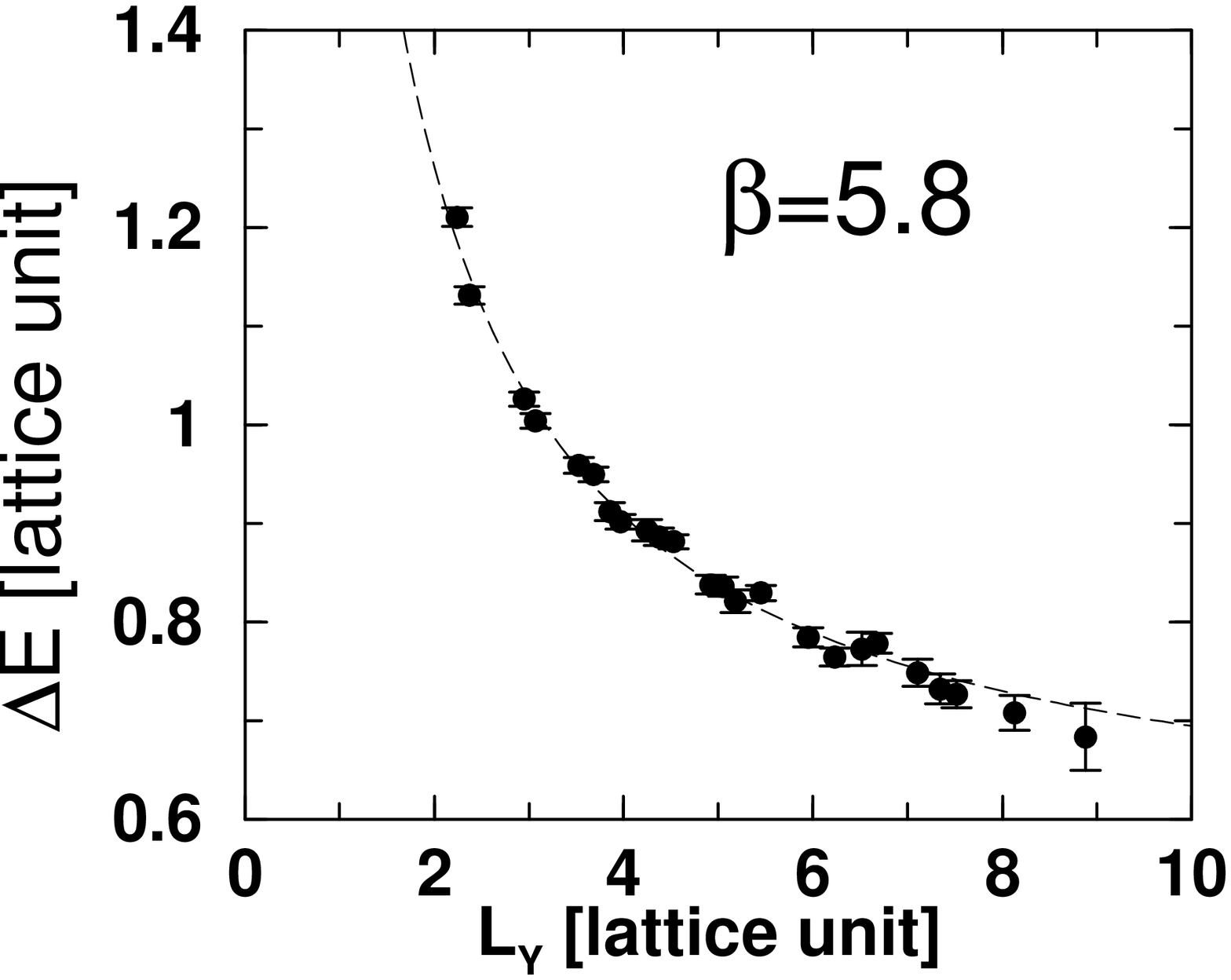}
\includegraphics[height=3.2cm]{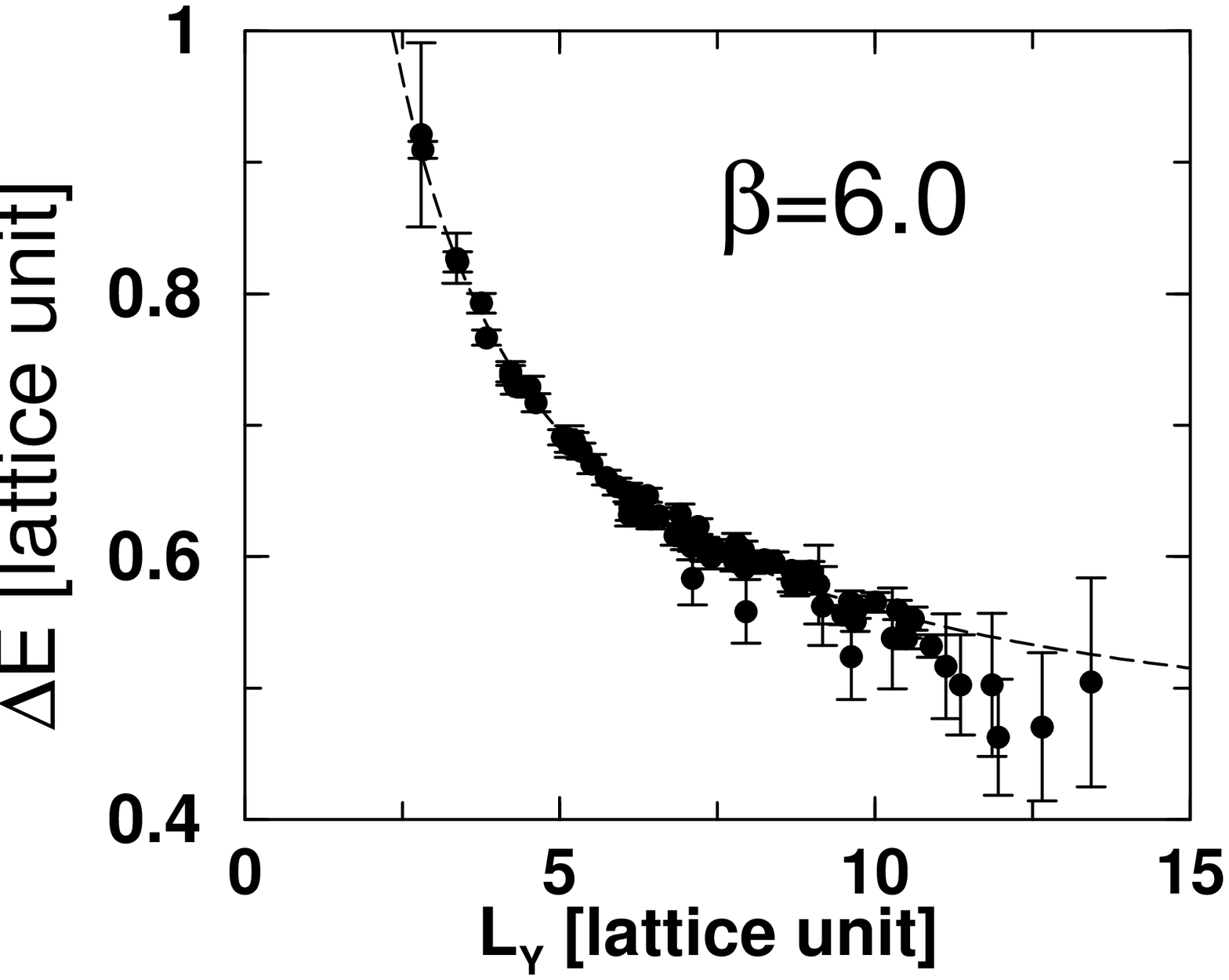}
\caption{
(a) The Mercedes form for the 3Q system. (b) \& (c) 
Lattice QCD results of the gluonic excitation energy 
$\Delta E_{\rm 3Q} \equiv V^{\rm e.s.}_{\rm 3Q}-V^{\rm g.s.}_{\rm 3Q}$
in the 3Q system plotted against the modified Y-length $L_{\overline{\rm Y}}$.
The dashed curve denotes the inverse Mercedes Ansatz. 
}
\end{center}
\end{figure}

The inverse Mercedes Ansatz indicates that the gluonic-excitation mode is realized 
as a complicated bulk excitation of the whole 3Q system. 

\subsection{Behind the Success of the Quark Model}

Here, we consider the connection between QCD and the quark model 
in terms of the gluonic excitation \cite{TS03,TS04,STI04,STOI04}. 
While QCD is described with quarks and gluons, 
the simple quark model successfully describes low-lying hadrons 
even without explicit gluonic modes.
In fact, the gluonic excitation seems invisible in low-lying hadron spectra, 
which is rather mysterious.

On this point, we find the gluonic-excitation energy to be about 1GeV or more, 
which is rather large compared with the excitation energies of the quark origin.
Therefore, the contribution of gluonic excitations 
is considered to be negligible and the dominant contribution is brought 
by quark dynamics such as the spin-orbit interaction for low-lying hadrons. 
Thus, the large gluonic-excitation energy of about 1GeV gives the physical reason for 
the invisible gluonic excitation in low-lying hadrons, 
which would play the key role for the success of the quark model 
without gluonic modes \cite{TS03,TS04,STI04,STOI04}.

In Fig.5, we present a possible scenario from QCD to the massive quark model 
in terms of color confinement and dynamical chiral-symmetry breaking.

\begin{figure}[h]
\begin{center}
\includegraphics[height=7.5cm]{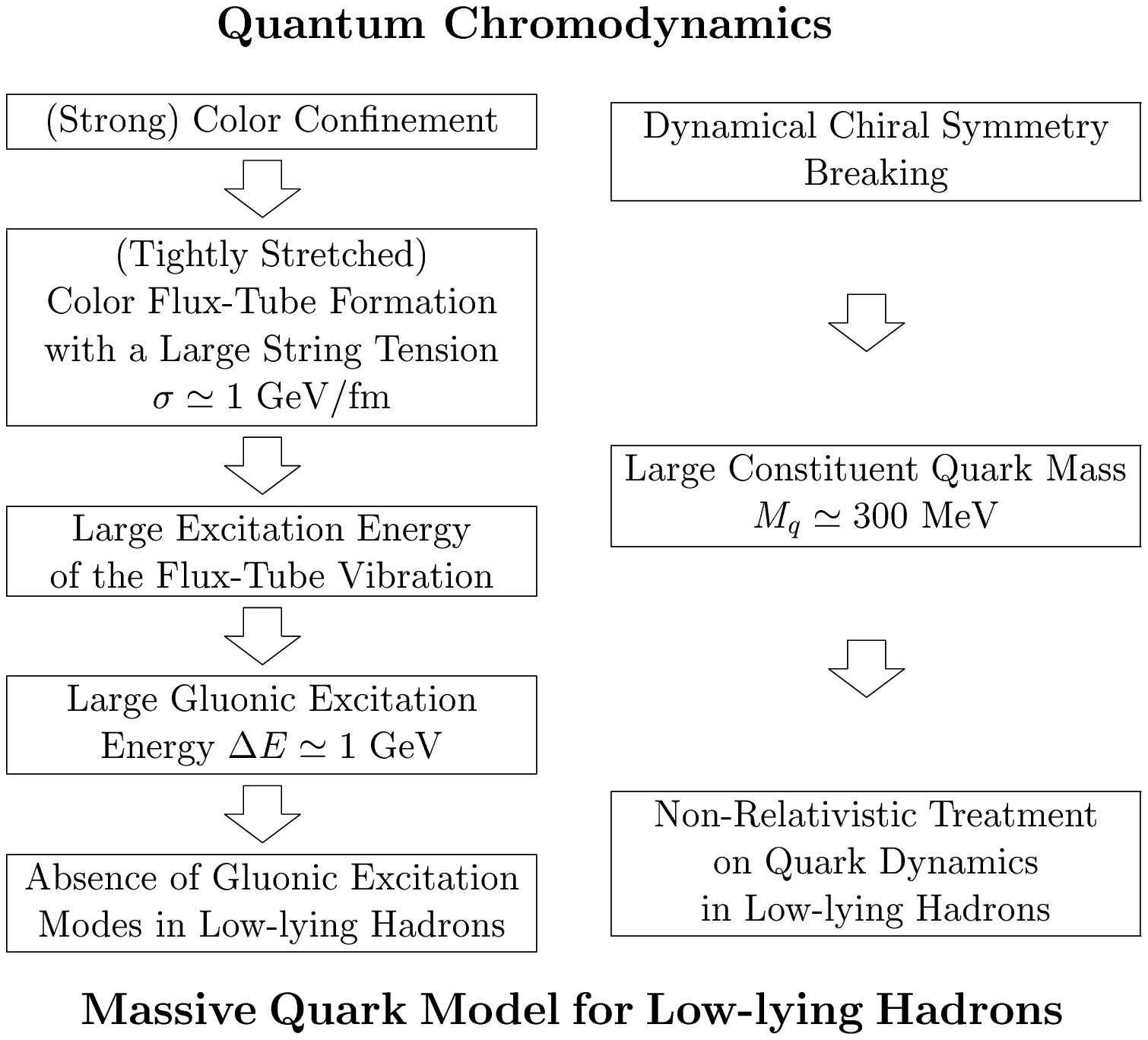}
\caption{A possible scenario from QCD to the quark model 
in terms of color confinement and dynamical chiral-symmetry breaking (DCSB).
DCSB leads to a large constituent quark mass of about 300 MeV, which enables the non-relativistic treatment 
for quark dynamics approximately. 
Color confinement results in the color flux-tube formation among quarks with a large string tension of $\sigma \simeq$ 1 GeV/fm.
In the flux-tube picture, the gluonic excitation is described as the flux-tube vibration, 
and its energy is expected to be large in the hadronic scale.
The large gluonic-excitation energy of about 1 GeV leads to 
the absence of the gluonic mode in low-lying hadrons, 
which plays the key role to the success of the quark model without gluonic excitation modes.
}
\end{center}
\vspace{-0.6cm}
\end{figure}

\newpage

\section{Lattice QCD Study for the Inter-Quark Interaction in Multi-Quark Systems}

In this section, we perform the first study of 
the inter-quark interaction in multi-quark systems in 
lattice QCD \cite{STOI04,OST04,OST04p,SOTI04,OST05}, 
and seek for the QCD-based quark-model Hamiltonian to describe multi-quark hadrons \cite{Z04}.
The quark-model Hamiltonian consists of the kinetic term and the potential term,
which is not known form QCD in multi-quark systems.

As for the potential at short distances, 
the perturbative one-gluon-exchange (OGE) potential would be appropriate, 
due to the asymptotic nature of QCD. 
For the long-range part, however, there appears the confinement potential as a typical 
non-perturbative property of QCD, and its form is highly nontrivial 
in the multi-quark system. 

In fact, 
to clarify the confinement force in multi-quark systems 
is one of the essential points for the 
construction of the QCD-based quark-model Hamiltonian.
Then, in this paper, we investigate the multi-quark potential in lattice QCD, 
with paying attention to the confinement force in multi-quark hadrons.

\subsection{The OGE Coulomb plus Multi-Y Ansatz}

We first consider the theoretical form of the multi-quark potential, 
since we will have to analyze the lattice QCD data by comparing them 
with some theoretical Ansatz.

By generalizing the lattice QCD result of the Y-Ansatz for the three-quark potential, 
we propose the one-gluon-exchange (OGE) Coulomb plus 
multi-Y Ansatz \cite{STOI04,OST04,OST04p,SOTI04,OST05},
\begin{eqnarray}
V=\frac{g^2}{4\pi}\sum_{i<j}\frac{T^a_i T^a_j}{|{\bf r}_i-{\bf r}_j|}+\sigma L_{\rm min}+C,
\label{VnQ}
\end{eqnarray}
for the potential form of the multi-quark system.
Here, the confinement potential is proportional to the minimal total length $L_{\rm min}$ 
of the color flux tube linking the quarks, which is multi-Y shaped in most cases.

In the following, we study the inter-quark interaction in multi-quark systems 
in lattice QCD, and compare the lattice QCD data with the theoretical form in Eq.(\ref{VnQ}).
Note here that the lattice QCD data are meaningful as primary data 
on the multi-quark system directly based on QCD, and do not depend on any theoretical Ansatz.

\subsection{Formalism of the Multi-Quark Wilson Loop}

Next, we formulate the multi-quark Wilson loop to obtain 
the multi-quark potential in lattice QCD \cite{STOI04,OST04,OST04p,SOTI04,OST05}.

Similar to the derivation of the Q$\rm\bar{Q}$ potential from the Wilson loop, 
the static multi-quark potential can be derived from 
the corresponding multi-quark Wilson loop.  
We construct the tetraquark Wilson loop $W_{\rm 4Q}$ and the pentaquark Wilson loop $W_{\rm 5Q}$ in a gauge invariant manner 
as shown in Figs.6(a) and (b), respectively. 

\begin{figure}[ht]
\centering
\includegraphics[scale=0.33,clip]{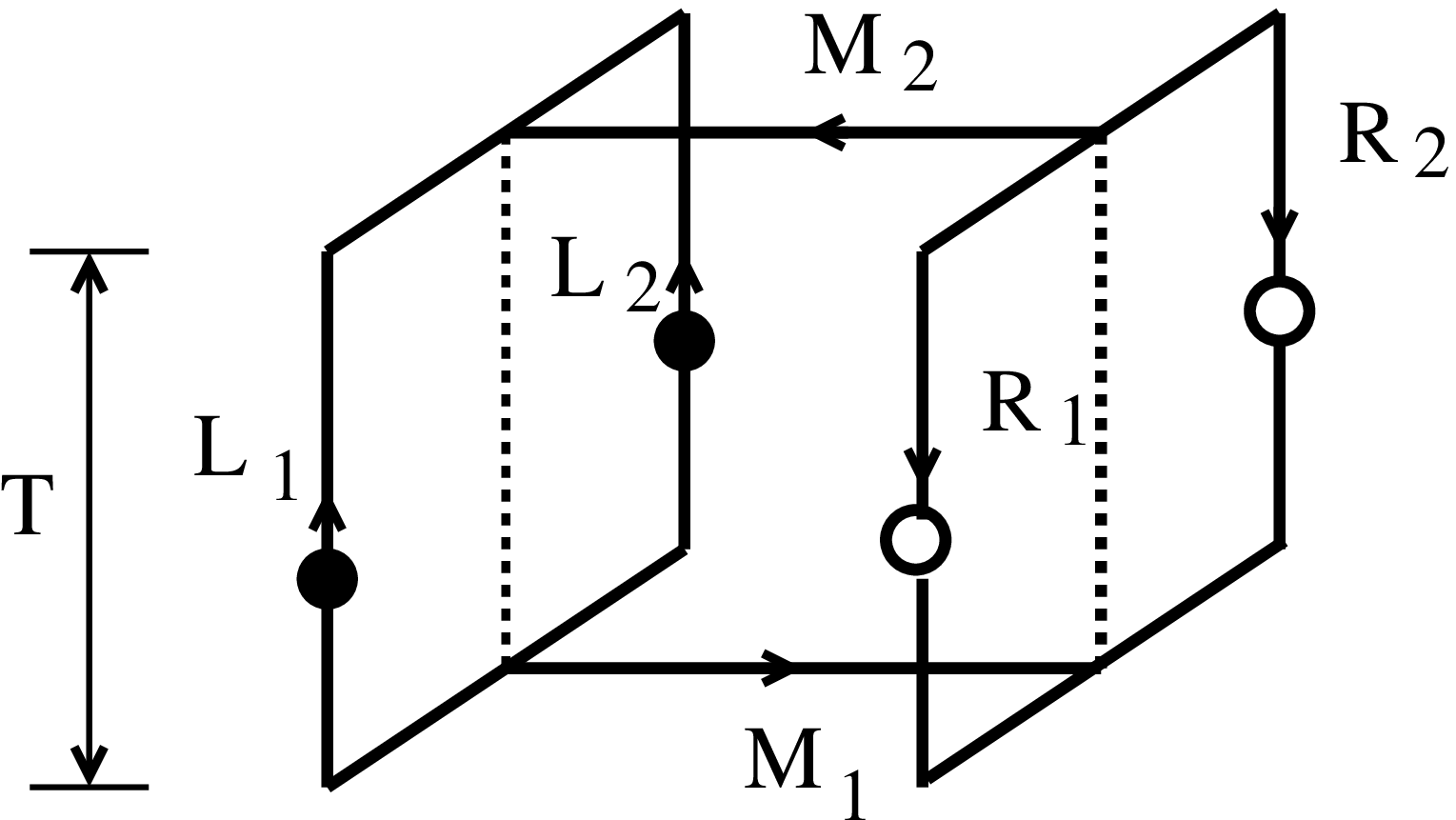}
\includegraphics[scale=0.33,clip]{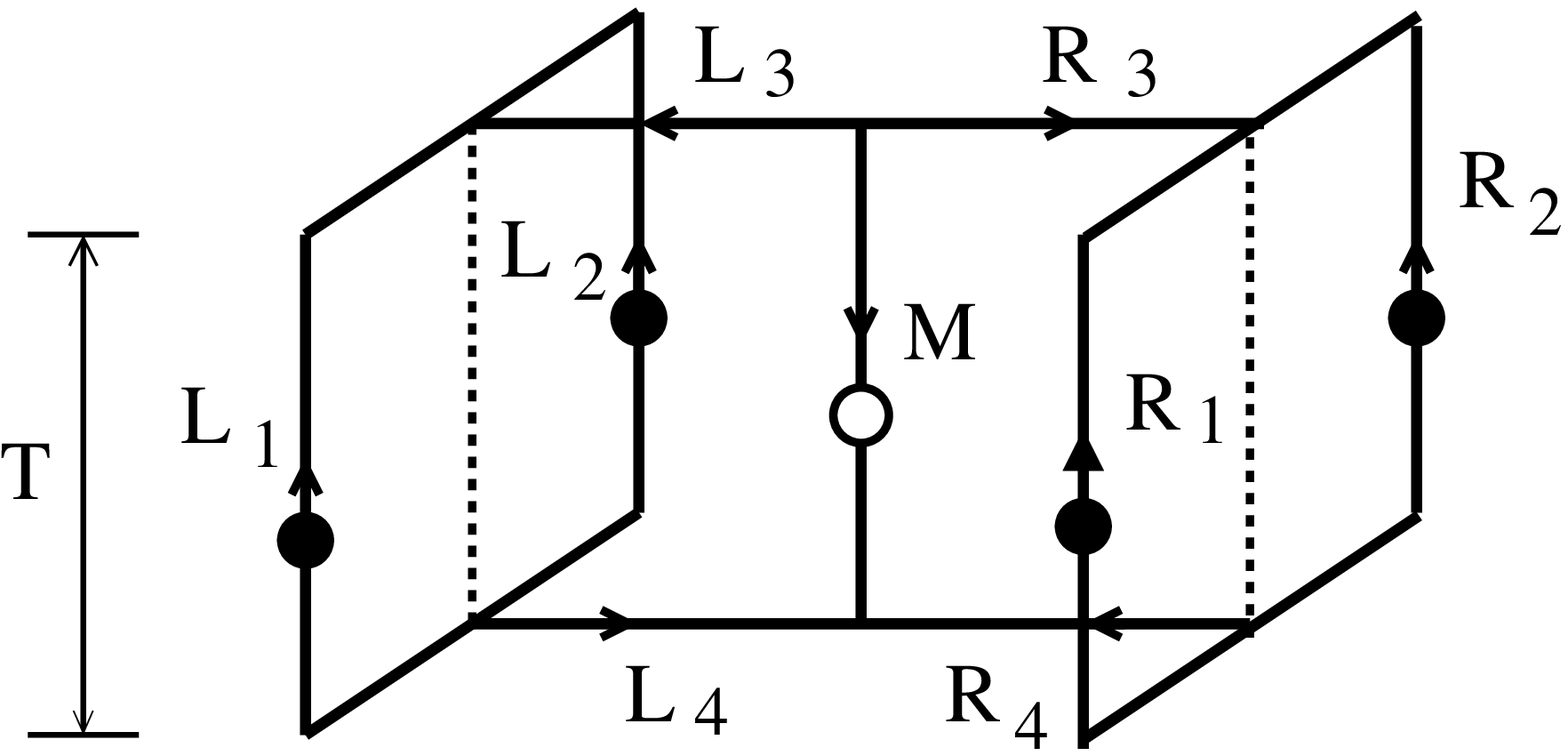}
\caption{(a) The tetraquark Wilson loop $W_{\rm 4Q}$. (b) The pentaquark Wilson loop $W_{\rm 5Q}$. The contours $M,M_i,R_j,L_j (i=1,2,j=3,4)$ are line-like and $R_j,L_j (j=1,2)$ are 
staple-like. The multi-quark Wilson loop physically means that 
a gauge-invariant multi-quark state is generated at $t=0$ and annihilated at $t=T$ with quarks being spatially fixed in ${\bf R}^3$ 
for $0<t<T$.}
\end{figure}

The tetraquark Wilson loop $W_{\rm 4Q}$ and the pentaquark Wilson loop $W_{\rm 5Q}$ are defined by 
\begin{eqnarray}
W_{\rm 4Q}&\equiv& \frac{1}{3}{\rm tr}(\tilde{M}_1 \tilde{R}_{12} \tilde{M}_2 \tilde{L}_{12}),\nonumber \\
W_{\rm 5Q}&\equiv& \frac1{3!}\epsilon^{abc}\epsilon^{a'b'c'}M^{aa'}(\tilde R_3\tilde R_{12}\tilde R_4)^{bb'}(\tilde L_3\tilde L_{12}\tilde L_4)^{cc'}, 
\end{eqnarray}
where $\tilde{M}$, $\tilde{M}_i$, $\tilde{L_j}$ and $\tilde{R_j}$ ($i$=1,2, $j$=1,2,3,4) are given by 
\begin{eqnarray}
\tilde{M}, \tilde{M}_i, \tilde{R_j}, \tilde{L_j}
\equiv 
P \exp{\{ig \int_{M, M_i,R_j,L_j}dx^\mu A_\mu (x)\}}\in \rm{SU(3)_c}.
\end{eqnarray}
Here, $\tilde{R}_{12}$ and $\tilde{L}_{12}$ are defined by
\begin{eqnarray}
\tilde{R}_{12}^{a'a} 
\equiv \frac{1}{2}\epsilon^{abc}\epsilon^{a'b'c'}
R_1^{bb'}R_2^{cc'},\quad 
\tilde{L}_{12}^{a'a} 
\equiv \frac{1}{2}\epsilon^{abc}\epsilon^{a'b'c'}
L_1^{bb'}L_2^{cc'}.
\end{eqnarray}
The multi-quark Wilson loop physically means that 
a gauge-invariant multi-quark state is generated at $t=0$ and annihilated at $t=T$ 
with quarks being spatially fixed in ${\bf R}^3$ for $0<t<T$.

The multi-quark potential is obtained from the vacuum expectation value of 
the multi-quark Wilson loop as
\begin{eqnarray}
V_{\rm 4Q}=-\lim_{T\rightarrow \infty} \frac1{T} {\rm ln} \langle W_{\rm 4Q}\rangle, 
\qquad
V_{\rm 5Q}=-\lim_{T\rightarrow \infty} \frac1{T} {\rm ln} \langle W_{\rm 5Q}\rangle.
\end{eqnarray}

\subsection{Lattice QCD Result of the Pentaquark Potential}

We perform the first study of the pentaquark potential $V_{\rm 5Q}$ in lattice QCD with ($\beta$=6.0, $16^3\times 32$)
for 56 different patterns of QQ-$\rm \bar Q$-QQ type pentaquark configurations, as shown in Fig.7.
As the conclusion, the lattice QCD data of  
$V_{\rm 5Q}$ are found to be well described by the OGE Coulomb plus multi-Y Ansatz, i.e., 
the sum of the OGE Coulomb term and the multi-Y-type 
linear term based on the flux-tube picture \cite{STOI04,OST04,OST04p,SOTI04,OST05}. 

\begin{figure}[h]
\begin{center}
\includegraphics[height=3.5cm]{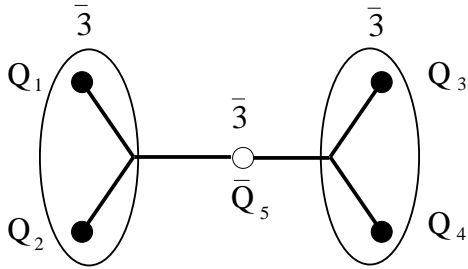} 
\caption{
A QQ-$\rm \bar Q$-QQ type pentaquark configuration.
In the 5Q system, 
$\rm (Q_1, Q_2)$ and $\rm (Q_3, Q_4)$ 
form $\bar {\bf 3}$ representation of SU(3) color, respectively.
The lattice QCD results indicate the multi-Y-shaped flux-tube formation 
in the QQ-$\rm \bar Q$-QQ system.
}
\end{center}
\end{figure}

We show in Fig.8 the lattice QCD results of the 5Q potential $V_{\rm 5Q}$ 
for symmetric planar 5Q configurations as shown in Fig.7, 
where each 5Q system is labeled by 
$d\equiv \overline{{\rm Q}_1{\rm Q}_2}/2$ and $h\equiv 
\overline{{\rm Q}_1{\rm Q}_3}$.

\begin{figure}[h]
\centering
\hspace{-0.2cm}
\includegraphics[height=3.6cm]{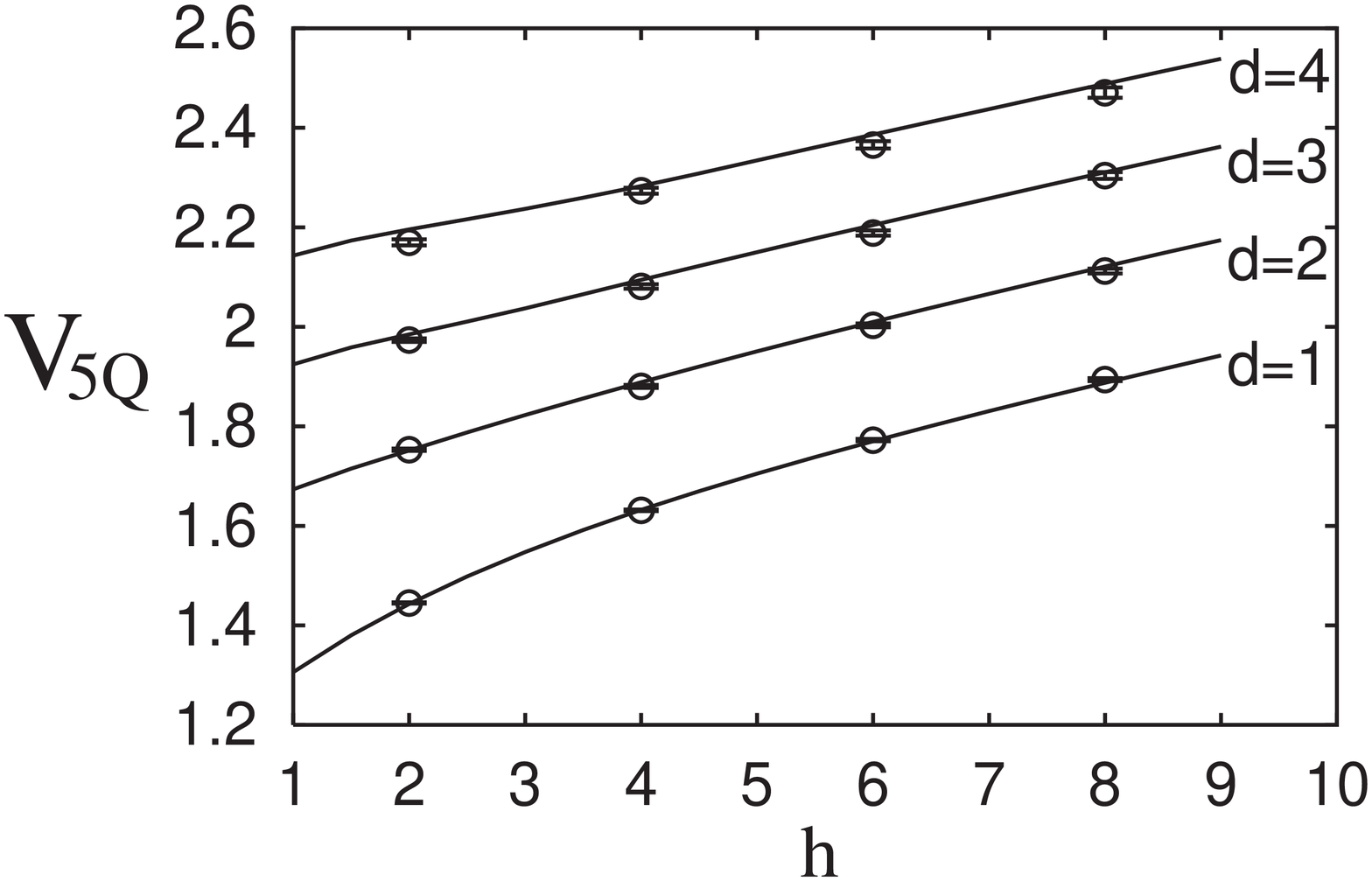}
\includegraphics[height=3.6cm]{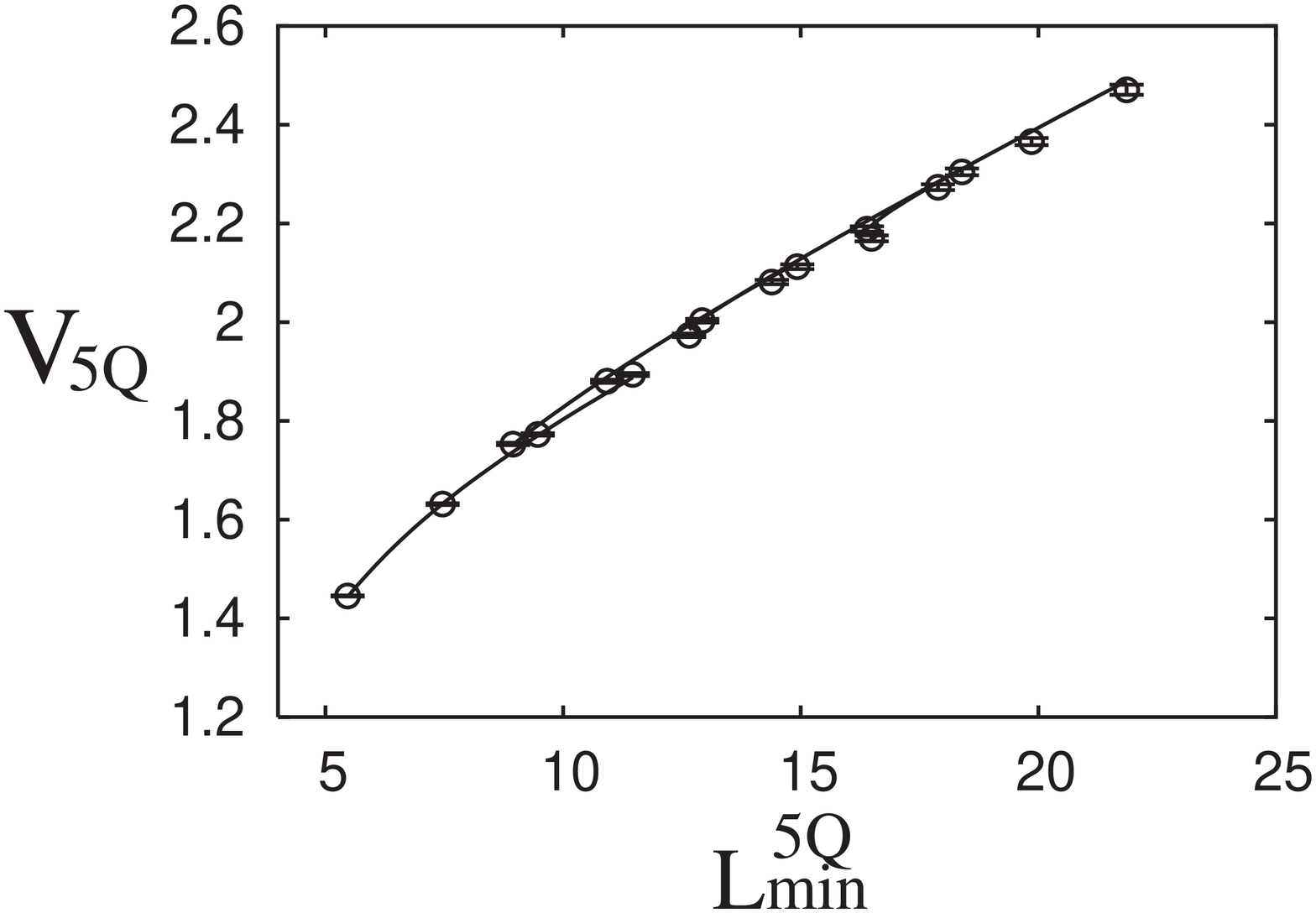}
\caption{Lattice QCD results of the pentaquark potential $V_{\rm 5Q}$ 
for symmetric planar 5Q configurations in the lattice unit:  
(a) $V_{\rm 5Q}$ v.s. $(d,h)$ and (b) $V_{\rm 5Q}$ v.s. $L_{\rm min}^{\rm 5Q}$.
The symbols denote the lattice QCD data, and the curves the theoretical form 
of the OGE plus multi-Y Ansatz.
}
\end{figure}

\noindent
In Fig.8, we add the theoretical curves of  
the OGE Coulomb plus multi-Y Ansatz, where the coefficients 
$(A_{\rm 5Q},\sigma_{\rm 5Q})$ are set to be 
$(A_{\rm 3Q},\sigma_{\rm 3Q})$ obtained from the 3Q potential \cite{TSNM02}.
(Note that there is no adjustable parameter in the theoretical Ansatz  
apart from an irrelevant constant.)
In Fig.8, one finds a good agreement between 
the lattice QCD data of $V_{\rm 5Q}$ and the theoretical curves of 
the OGE Coulomb plus multi-Y Ansatz.

In this way, the pentaquark potential $V_{\rm 5Q}$ is found to be well described 
by the OGE Coulomb plus multi-Y Ansatz as \cite{STOI04,OST04,OST04p,SOTI04} 
\begin{eqnarray}
V_{\rm 5Q}=&-&A_{\rm 5Q}\{ ( \frac1{r_{12}}  + \frac1{r_{34}}) 
+\frac12(\frac1{r_{15}} +\frac1{r_{25}} +\frac1{r_{35}} +\frac1{r_{45}}) \nonumber \\
&+&\frac14(\frac1{r_{13}} +\frac1{r_{14}} +\frac1{r_{23}} +\frac1{r_{24}}) \}
+\sigma_{\rm 5Q}L_{\rm min}^{\rm 5Q}+C_{\rm 5Q}, 
\label{V5Q}
\end{eqnarray}
where $r_{ij}$ 
is the distance between ${\rm Q}_i$ and ${\rm Q}_j$. 
Here, $L_{\rm min}^{\rm 5Q}$ is the minimal total length of the 
flux tube, which is multi-Y-shaped as shown in Fig.7.
This lattice result supports the flux-tube picture for the 5Q system.

\subsection{Tetraquark Potential and Flip-Flop in Lattice QCD}

For about 200 different patterns of QQ-${\rm \bar{Q}\bar{Q}}$ configurations, 
we perform the detailed study of the tetraquark potential $V_{\rm 4Q}$ in lattice QCD with $\beta$=6.0, $16^3\times 32$, 
and find the following results \cite{STOI04,OST04p,OST05}.
\begin{enumerate}
\item
When QQ and $\rm \bar Q \bar Q$ are well separated,  
the 4Q potential $V_{\rm 4Q}$ is well described by the OGE Coulomb plus multi-Y Ansatz,
which indicates the multi-Y-shaped flux-tube formation as shown in Fig.9(a).
\item
When the nearest quark and antiquark pair is spatially close, 
the 4Q potential $V_{\rm 4Q}$ is well described by the sum of two Q$\bar {\rm Q}$ potentials, 
which indicates a ``two-meson" state as shown in Fig.9(b).
\end{enumerate}

\begin{figure}[h]
\begin{center}
\includegraphics[height=3.5cm]{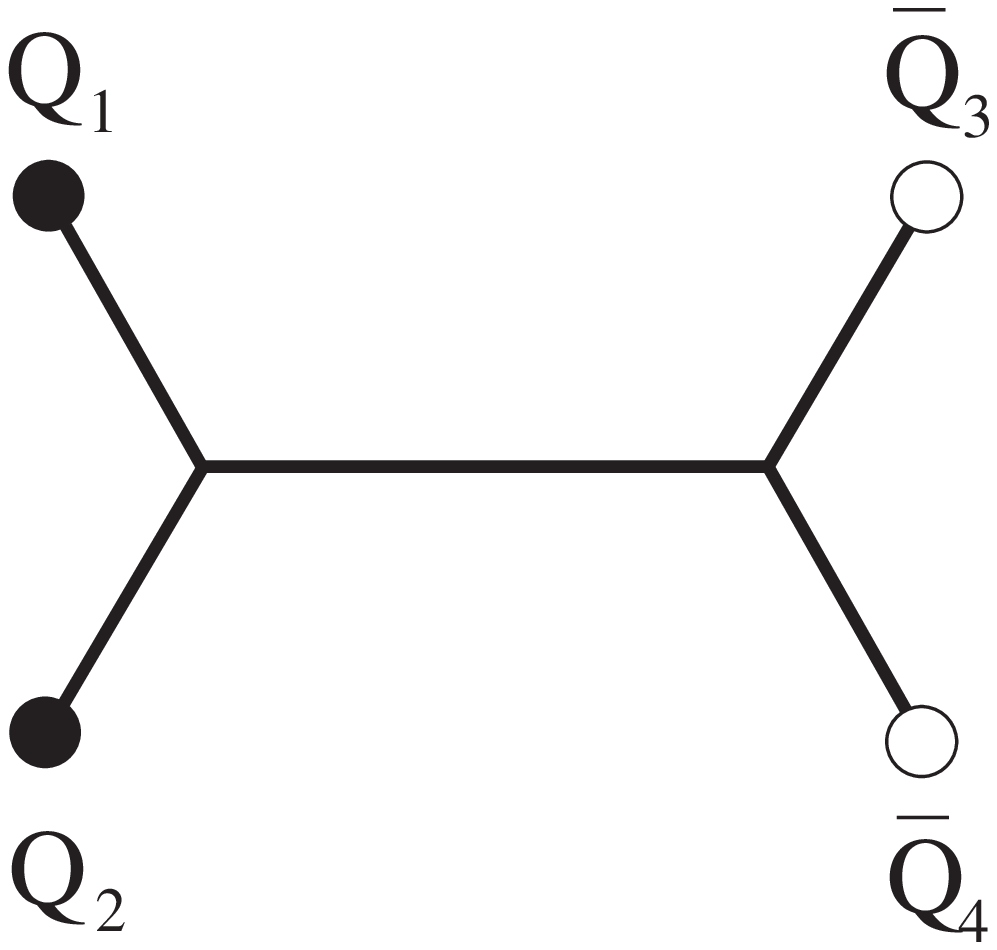}
\hspace{1.5cm}
\includegraphics[height=3.2cm]{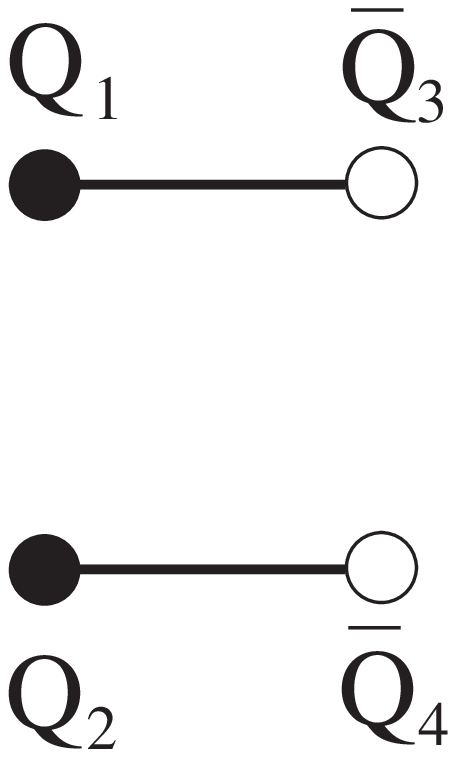} 
\caption{
(a) A connected tetraquark (QQ-$\rm \bar Q\bar Q$) configuration 
and (b) A ``two-meson" configuration.
The lattice QCD results indicate the multi-Y-shaped flux-tube formation 
for the connected 4Q system.
}
\end{center}
\end{figure}

We show in Fig.10 the lattice QCD results of the 4Q potential $V_{\rm 4Q}$ 
for symmetric planar 4Q configurations as shown in Fig.9, 
where each 4Q system is labeled by 
$d\equiv \overline{{\rm Q}_1{\rm Q}_2}/2$ and $h\equiv 
\overline{{\rm Q}_1{\rm Q}_3}$.

\begin{figure}[h]
\centering
\includegraphics[height=3.8cm,clip]{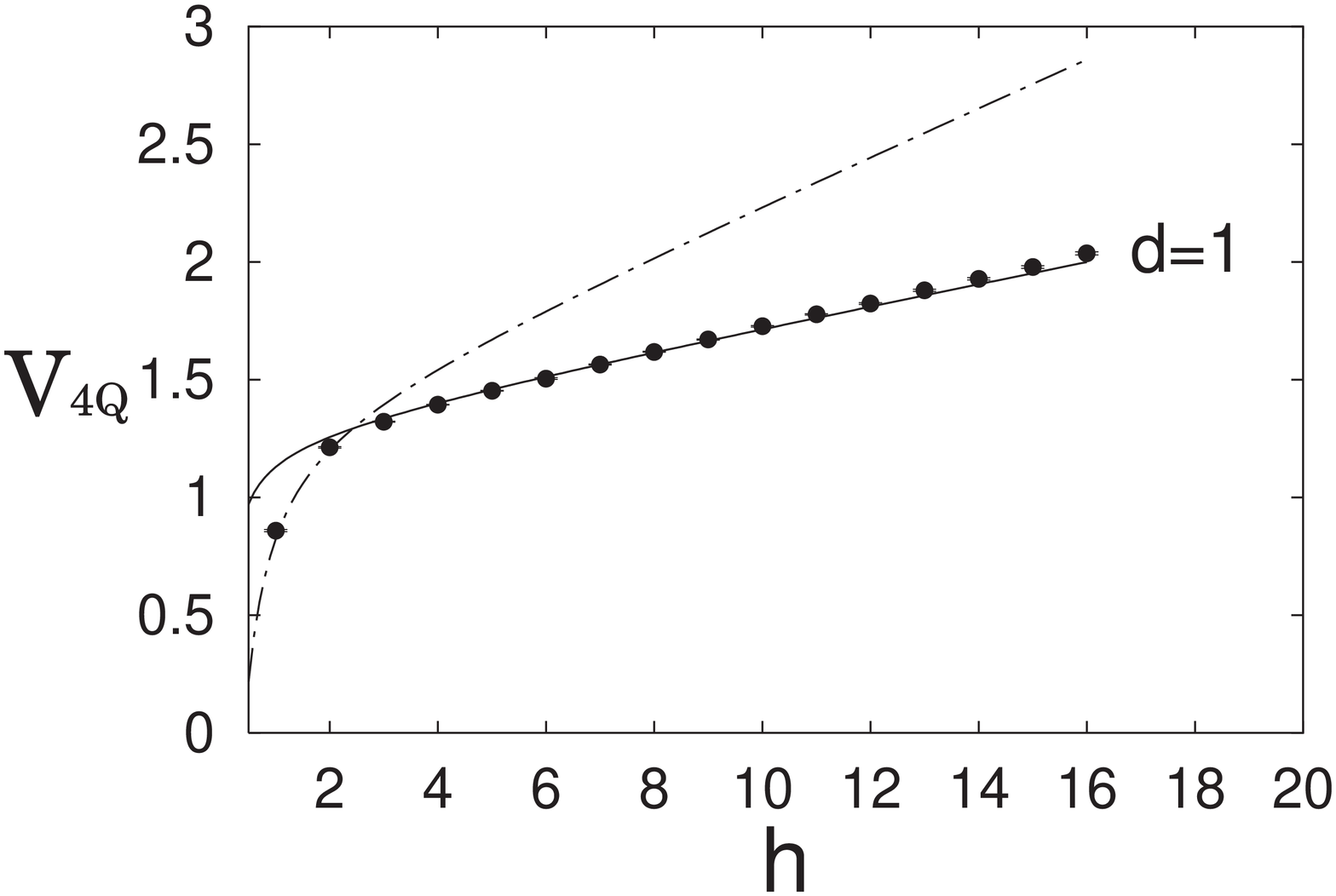}
\hspace{0.1cm}
\includegraphics[height=3.8cm,clip]{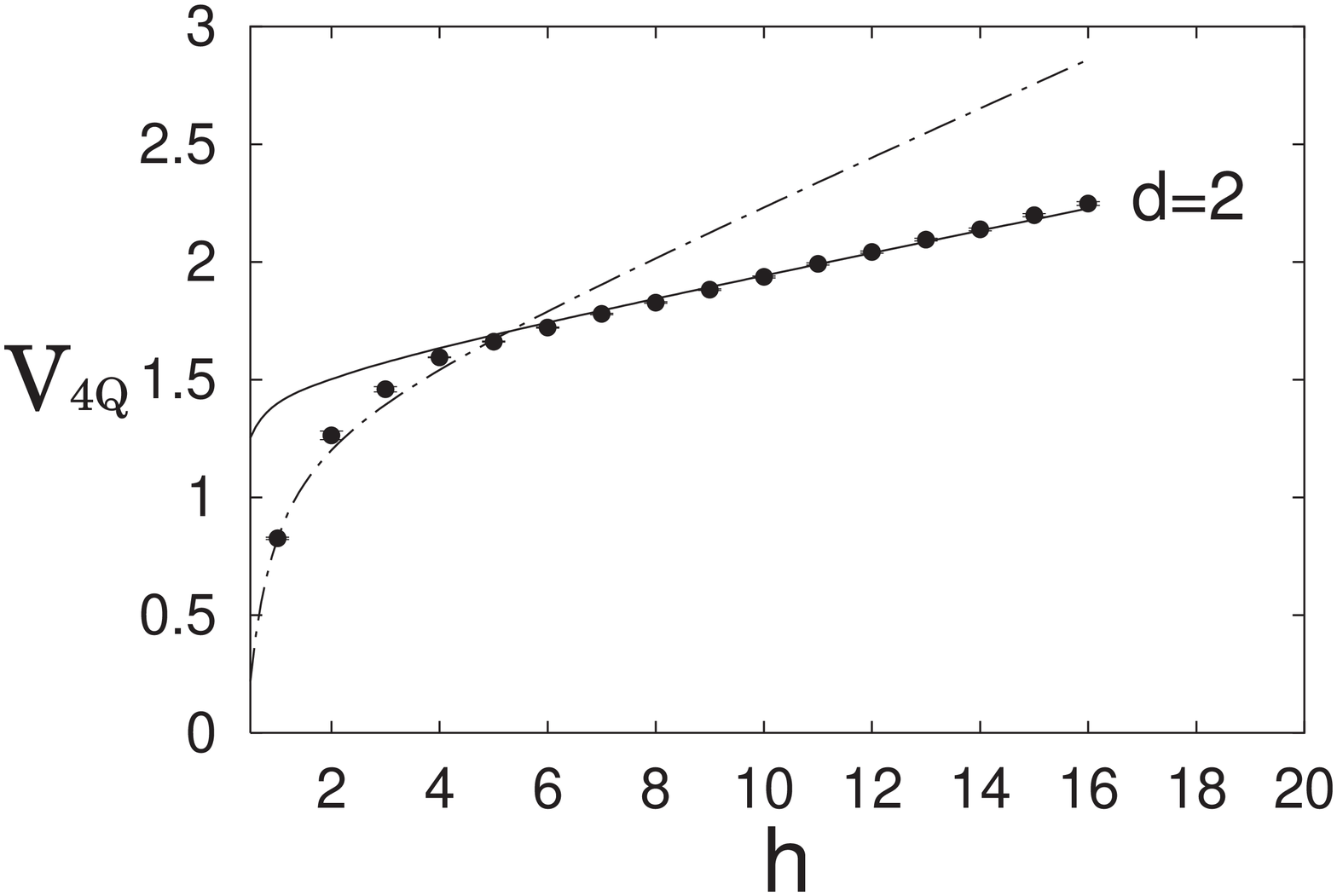}
\caption{
Lattice QCD results of the tetraquark potential $V_{\rm 4Q}$ 
for symmetric planar 4Q configurations in the lattice unit.
The symbols denote the lattice QCD data. 
The solid curve denotes the OGE plus multi-Y Ansatz, 
and the dotted-dashed curve the two-meson Ansatz.}
\end{figure}

For large value of $h$ compared with $d$, the lattice data seem to coincide with 
the solid curve of the OGE Coulomb plus multi-Y Ansatz,
\begin{eqnarray}
V_{\rm 4Q}
&=&-A_{\rm 4Q}\left\{ \left( \frac{1}{r_{12}}+\frac{1}{r_{34}} \right)+\frac1{2}\left( \frac{1}{r_{13}}+\frac{1}{r_{14}}+\frac{1}{r_{23}}+\frac{1}{r_{24}} \right) \right \} \nonumber \\
&&+\sigma_{\rm 4Q}L_{\rm min}^{\rm 4Q}+C_{\rm 4Q},
\label{V4Q}
\end{eqnarray}
where $L_{\rm min}^{\rm 4Q}$ is the minimal total length of 
the flux tube, which is multi-Y-shaped as shown in Fig.9(a). 
Here, the coefficients $(A_{\rm 4Q},\sigma_{\rm 4Q})$ are set to be 
$(A_{\rm 3Q},\sigma_{\rm 3Q})$ obtained from the 3Q potential \cite{TSNM02}.

For small $h$, 
the lattice data tend to agree with 
the dotted-dashed curve of the ``two-meson" Ansatz, 
where the 4Q potential is described 
by the sum of two Q$\bar {\rm Q}$ potentials as 
$V_{\rm Q\bar Q}(r_{13})+V_{\rm Q\bar Q}(r_{24})=2V_{\rm Q\bar Q}(h)$. 

Thus, the tetraquark potential $V_{\rm 4Q}$ is found to take 
the smaller energy of the connected 4Q state or the two-meson state.
In other words, we observe a clear lattice QCD evidence 
of the ``flip-flop", i.e., the flux-tube recombination 
between the connected 4Q state and the two-meson state.
This lattice result also supports the flux-tube picture for the 4Q system.

\subsection{Proper Quark-Model Hamiltonian for Multi-Quarks}

From a series of our lattice QCD studies \cite{TMNS01,TSNM02,TS03,TS04,STI04,STOI04,OST04,OST04p,SOTI04,OST05} 
on the inter-quark potentials, 
the inter-quark potential is clarified to consist of 
the one-gluon-exchange (OGE) Coulomb part  
and the flux-tube-type linear confinement part 
in Q$\rm \bar Q$-mesons, 3Q-baryons and multi-quark (4Q, 5Q) hadrons. 

Furthermore, from the comparison among 
the $\rm Q\bar Q$, 3Q, 4Q and 5Q potentials in lattice QCD,
we find the universality of the string tension $\sigma$,
\begin{eqnarray}
\sigma_{\rm Q\bar{\rm Q}}\simeq \sigma_{\rm 3Q} \simeq \sigma_{\rm 4Q} 
\simeq \sigma_{\rm 5Q},
\end{eqnarray}
and the OGE result of the Coulomb coefficient $A$ as 
\begin{eqnarray}
\frac1{2}A_{\rm Q\bar{\rm Q}}\simeq A_{\rm 3Q}
\simeq A_{\rm 4Q}\simeq A_{\rm 5Q}
\end{eqnarray}
in Eqs.(\ref{VQQ}), (\ref{V3Q}), (\ref{V5Q}) and (\ref{V4Q}).

Here, the OGE Coulomb term is considered to originate 
from the OGE process, which plays the dominant role at short distances, 
where perturbative QCD is applicable.
The flux-tube-type linear confinement would 
be physically interpreted by the flux-tube picture, where 
quarks and antiquarks are linked by the one-dimensional squeezed color-electric flux tube  
with the string tension $\sigma$. 

To conclude, the inter-quark interaction would 
be generally described by the sum of the short-distance two-body OGE part  
and the long-distance flux-tube-type linear confinement part 
with the universal string tension 
$\sigma \simeq 0.89$ GeV/fm.

Thus, based on the lattice QCD results, we propose 
the proper quark-model Hamiltonian $\hat H$ for multi-quark hadrons as 
\begin{eqnarray}
\hat H=\sum_{i} \sqrt{\hat {\bf p}_i^2+M_i^2}+\sum_{i<j}V_{\rm OGE}^{ij}
+\sigma L_{\rm min},
\end{eqnarray}
where $L_{\rm min}$ is the minimal total length of the flux tube linking quarks. 
$V_{\rm OGE}^{ij}$ denotes the OGE potential between $i$th and $j$th quarks,
which becomes the OGE Coulomb potential in Eq.(\ref{VnQ}) in the static case.
$M_i$ denotes the constituent quark mass.
The semi-relativistic treatment would be necessary for light quark systems.

It is desired to investigate various properties of multi-quark hadrons 
with this QCD-based quark model Hamiltonian $\hat H$.

\vspace{0.5cm}

\noindent
{\bf \large Acknowledgements}: 
H.S. sincerely thanks Profs. M.~Chabab and M.~Lazrek, 
and other organizers of ICHEMP'05 for their warm hospitality in Marrakech, Morocco.
The lattice QCD Monte Carlo calculations have been performed on NEC-SX5 at Osaka University 
and HITACHI-SR8000 at KEK.


\begin{thebibliography}{99}
\bibitem{N66} Y.~Nambu, in {\it Preludes in Theoretical Physics}, 
(North-Holland, 1966).
\bibitem{HN65} M.Y.~Han and Y.~Nambu, Phys. Rev. {\bf 139} (1965) B1006. 
\bibitem{GWP73}
D.J.~Gross and F.~Wilczek, Phys. Rev. Lett. {\bf 30} (1973) 1343; \\
H.D.~Politzer, Phys. Rev. Lett. {\bf 30} (1973) 1346.
\bibitem{N6970} Y.~Nambu, in {\it Symmetries and Quark Models} (Wayne State University, 1969); 
{\it Lecture Notes at the Copenhagen Symposium} (1970).
\bibitem{N74} Y.~Nambu, Phys. Rev. {\bf D10} (1974) 4262.
\bibitem{conf2003} For instance, articles in 
{\it Color Confinement and Hadrons in Quantum Chromodynamics}, 
edited by H.~Suganuma {\it et al.} (World Scientific, 2004).
\bibitem{NJL61} Y.~Nambu and G.~Jona-Lasinio, 
Phys. Rev. {\bf 122} (1961) 345; {\it ibid.} {\bf 124} (1961) 246.
\bibitem{C7980} M.~Creutz, 
Phys. Rev. Lett. {\bf 43} (1979) 553; Phys. Rev. {\bf D21} (1980) 2308.
\bibitem{R97} 
H.J.~Rothe, {\it Lattice Gauge Theories}, 2nd edition (World Scientific, 1997).
\bibitem{TMNS99}
T.T.~Takahashi, H.~Matsufuru, Y.~Nemoto and H.~Suganuma, 
in {\it Dynamics of Gauge Fields}, Tokyo, Dec. 1999, edited by A.~Chodos {\it et al.}, 
(Universal Academy Press, 2000) 179; 
H.~Suganuma, Y.~Nemoto, H.~Matsufuru and T.T.~Takahashi, Nucl. Phys. {\bf A680}, 159 (2000).
\bibitem{TMNS01} T.T.~Takahashi, H.~Matsufuru, Y.~Nemoto and H.~Suganuma, 
Phys.~Rev.~Lett. {\bf 86} (2001) 18.
\bibitem{TSNM02}T.T.~Takahashi, H.~Suganuma, Y.~Nemoto and H.~Matsufuru, 
Phys.~Rev.~{\bf D65} (2002) 114509.
\bibitem{TS03} T.T.~Takahashi and H.~Suganuma, Phys. Rev. Lett. {\bf 90} (2003) 182001.
\bibitem{TS04} T.T.~Takahashi and H.~Suganuma, Phys. Rev. {\bf D70} (2004) 074506.
\bibitem{STI04} H.~Suganuma, T.T.~Takahashi and H.~Ichie, 
in {\it Color Confinement and Hadrons in Quantum Chromodynamics}, 
(World Scientific, 2004) p.249.
\bibitem{STOI04}
H.~Suganuma, T.~T.~Takahashi, F.~Okiharu and H.~Ichie, 
in {\it QCD Down Under}, March 2004, Adelaide, 
Nucl. Phys. {\bf B} (Proc. Suppl.) {\bf 141} (2005) 92;
in {\it Quark Confinement and Hadron Spectrum VI}, 
September 2004, Sardinia, Italy, 
AIP Conference Proceedings {\bf CP756} (2005) 123.
\bibitem{OST04}
F.~Okiharu, H.~Suganuma and T.~T.~Takahashi, 
Phys.~Rev.~Lett. {\bf 94} (2005) 192001.
\bibitem{OST04p}
F.~Okiharu, H.~Suganuma, T.~T.~Takahashi, 
in {\it Pentaquark04}, July 2004, SPring-8, Japan (World Scientific, 2005) 339.
\bibitem{SOTI04}
H. Suganuma, F. Okiharu, T.T. Takahashi and H. Ichie, Nucl. Phys. {\bf A755} (2005) 399; 
H.~Suganuma, H.~Ichie, F.~Okiharu and T.T.~Takahashi, in {\it Pentaquark04}, July 2004, SPring-8, Japan
(World Scientific, 2005) 414. 
\bibitem{OST05}
F.~Okiharu, H.~Suganuma and T.~T.~Takahashi, Phys. Rev. {\bf D72} (2005) 014505.
\bibitem{KS75CKP83} J.~Kogut and L.~Susskind, Phys. Rev. {\bf D11} (1975) 395;
J.~Carlson, J.~Kogut and V.~Pandharipande, Phys. Rev. {\bf D27} (1983) 233; {\it ibid.} {\bf D28} (1983) 2807.
\bibitem{FRS91} M.~Fable de la Ripelle and Yu. A.~Simonov, Ann. Phys. {\bf 212} (1991) 235. 
\bibitem{BPV95} N.~Brambilla, G.M.~Prosperi and A.~Vairo, Phys. Lett. {\bf B362} (1995) 113.
\bibitem{IBSS03} H.~Ichie, V.~Bornyakov, T.~Streuer and G.~Schierholz, 
Nucl. Phys. {\bf A721}, 899 (2003); Nucl. Phys. {\bf B} (Proc.Suppl.) {\bf 119} (2003) 751. 
\bibitem{KS03} D.S. Kuzmenko and Yu.A. Simonov, Phys. Atom. Nucl. {\bf 66} (2003) 950.
\bibitem{C0405} J.M.~Cornwall, Phys. Rev. {\bf D69}, 065013 (2004); {\it ibid.} {\bf D71} (2005) 056002.
\bibitem{BS04} P.O.~Bowman and A.P.~Szczepaniak, Phys. Rev. {\bf D70} (2004) 016002.
\bibitem{JKM03} K.J.~Juge, J.~Kuti and C.~Morningstar, Phys. Rev. Lett. {\bf 90} (2003) 161601.
\bibitem{Z04} For a recent review, 
S.~L.~Zhu, Int. J. Mod. Phys. {\bf A19} (2004) 3439. 

\end{thebibliography}
\end{document}